\documentstyle[12pt,aasms4]{article}

\begin{document}

\title{Correlations between Low-Frequency QPOs and Spectral Parameters in
XTE~J1550--564 and GRO~J1655--40}
\author{Gregory J. Sobczak}
\affil{Astronomy Dept., Harvard University, 60 Garden St., Cambridge, MA
02138; gsobczak@cfa.harvard.edu}
\author{Jeffrey E. McClintock}
\affil{Harvard-Smithsonian Center for Astrophysics, 60 Garden St., Cambridge, MA
02138; jem@cfa.harvard.edu}
\author{Ronald A. Remillard, Wei Cui, Alan M. Levine, and Edward H. Morgan}
\affil{Center for Space Research, MIT, Cambridge, MA 02139; rr@space.mit.edu, 
cui@space.mit.edu, aml@space.mit.edu, ehm@space.mit.edu}
\author{Jerome A. Orosz}
\affil{Sterrekundig Instituut, Universiteit Utrecht, Postbus 80.000, 3508 TA Utrecht,
The Netherlands; J.A.Orosz@astro.uu.nl}
\author{Charles D. Bailyn}
\affil{Department of Astronomy, Yale University, P. O. Box 208101, New Haven, CT
06520; bailyn@astro.yale.edu}

\begin{abstract} 

Utilizing observations obtained with the {\it Rossi X-ray Timing Explorer}, we examine
correlations between the properties of 0.08--22~Hz variable frequency QPOs and the
X-ray spectral parameters for the black hole candidates XTE~J1550--564 and
GRO~J1655--40.  The spectra were fitted to a model including a multi-temperature
blackbody disk and a power-law component.  We find that the QPO frequency and
amplitude are well correlated with the spectral parameters, although the correlations
found for XTE~J1550--564 are generally opposite to those for GRO~J1655--40.  There is
one exeption: Both sources exhibit a general increase in the QPO frequency as the disk
flux increases (or as the mass accretion rate through the disk increases).  In
addition, these QPOs are observed only when the power-law component contributes more
than 20\% of the 2--20~keV flux, which indicates that both the disk and the power-law
components are linked to the QPO phenomenon.  

\end{abstract}


\section{Introduction}

GRO~J1655--40 is a black hole X-ray nova (BHXN) with a dynamically-determined primary
mass of $\sim7~M_{\odot}$ (Orosz \& Bailyn 1997; Shahbaz et al.~1999), which exceeds
the maximum mass of a neutron star (Rhoades \& Ruffini 1974; Kalogera \& Baym 1996). 
The source displays superluminal radio jets and is at a distance of $3.2\pm0.2$~kpc
(Hjellming \& Rupen 1995).  GRO~J1655--40 was discovered with BATSE on 1994 July~27,
and in early 1996 it entered a very low or quiescent state.  However, a second
outburst commenced on 1996 April~25 and continued for 16~months.  During this second
outburst, there was an intensive campaign of observations with the {\it Rossi X-ray
Timing Explorer} (RXTE); this paper is based on extensive spectral and timing results
derived from these 52 1996-97 observations (Sobczak et al.~1999a; Remillard et
al.~1999a; see also Mendez et al.~1998).

XTE~J1550--564 is an X-ray nova (aka soft X-ray transient) and a black hole candidate,
which was discovered on 1998 September~6 (Smith et al.~1998) .  Two weeks later it
reached a peak intensity of 6.8~Crab at 2--10~keV to become the brightest X-ray nova
yet observed with RXTE. The source has been observed in the very high, high/soft, and
intermediate canonical outburst states of BHXN (Sobczak et al.~1999bc; Cui et
al.~1999).  There is both a radio counterpart (Campbell-Wilson et al.~1998) and an
optical counterpart, which has been studied extensively in outburst (Jain et al.~1999;
Sanchez-Fernandez et al.~1999).  The mass of the compact primary is unknown.  The data
presented herein are based on 169 pointed observations with RXTE, which represent all
of the pointed observations of XTE~J1550--564 during Gain Epoch~3 (i.e. before 16:30
(UT) on 1999 March~22).  

Both GRO~J1655--40 and XTE~J1550--564 are of intense current interest because they
display a variety quasi-periodic X-ray oscillations (QPOs) which can be used to probe
the accretion process around black holes.  It is well known that the temporal
characteristics of black hole candidates are strongly correlated with the spectral
state (see van der Klis 1995, and references therein).  QPOs are fundamental
properties of the accretion flow and in a few sources the rms amplitude can exceed
15\% of the mean X-ray flux (see \S4).  Consequently, simultaneous studies of the
temporal and spectral properties of BHXN are a promising avenue for determining the
origin of QPOs in these sources.  

GRO~J1655--40 exhibits four types of QPOs between 0.1~Hz and 300~Hz (Remillard et
al.~1999a), three of which have relatively stable central frequencies.  A fourth QPO
with a central frequency that varies from 14--22~Hz is the only one of interest here. 
Similarly, the power spectra of XTE~J1550--564 exhibit two types of QPOs: a
high-frequency QPO ($\sim200$~Hz) and a variable 0.08--18~Hz QPO (Remillard et
al.~1999b; Sobczak et al.~1999b; Cui et al.~1999).  Again, we are interested only in
this latter QPO.  Our focus in this paper is on the correlations between the frequency
and amplitude of the 0.08--18~Hz and 14--22~Hz QPOs and the spectral parameters for
these two sources.  We explore the implications of these correlations for the origin
of these QPOs and compare our results with several QPO models.

\section{Observations and Analysis}

The timing and spectral data were obtained using the Proportional Counter Array (PCA;
Jahoda et al.~1996) onboard the {\it Rossi X-ray Timing Explorer} (RXTE).  The PCA
consists of five xenon-filled detector units (PCUs) with a total effective area of
$\sim$~6200~cm$^{-2}$ at 5~keV.  The PCA is sensitive in the range 2--60~keV, the
energy resolution is $\sim$17\% at 5~keV, and the time resolution capability is
1~$\mu$sec.  

The PCA spectral data in the energy range 2.5--20~keV were fit to the widely used
model consisting of a multi-temperature blackbody accretion disk plus power-law
(Tanaka \& Lewin 1995, and references therein).  The spectral parameters include the
temperature and radius of the inner accretion disk and the power-law photon index; the
power-law flux and the disk flux were derived from the fitted parameters.  
See Sobczak et al.~(1999abc) for additional information regarding spectral fitting
methods.  The fitted temperature and radius of the inner accretion disk presented here
are the observed color temperature and radius of the inner disk.  Since the disk
emission is likely to be affected by spectral hardening due to electron scattering
(Shakura \& Sunyaev 1973), the physical interpretation of these parameters remains
uncertain.  

An X-ray timing analysis was conducted by computing the 2--30~keV power spectrum for
each PCA observation.  The contribution from counting statistical noise was subtracted
and the power spectra were corrected for dead-time effects as described by Zhang et
al.~(1995) and Morgan, Remillard, \& Greiner (1997).  A chi-squared minimization
technique was used to derive the central frequency and width of each QPO.  The QPO
features were fit with Lorentzian functions, while the power continuum in the vicinity
of the QPO feature was modeled with a power-law function.  Triple QPO features
exhibiting a fundmental as well as sub- and first-harmonics were common (see Fig.~1a).
We identify the central QPO feature with the largest rms amplitude as the fundamental.
The integrated fractional rms amplitude of the QPO is the square root of the
integrated power in the QPO feature, expressed as a fraction of the mean count rate. 
The amplitudes of weaker QPOs at harmonic frequencies have been excluded from the QPO
amplitude discussed here.  See Remillard et al.~(1999ab) and for further information
regarding the timing analysis of the data for GRO~J1655--40 and XTE~J1550--564.  QPOs
observed during the rising phase of XTE~J1550--564 (observations 1--14 in Table~1)
were first reported by Cui et al.~(1999) and details of the timing analysis for those
observations can be found therein.  

For GRO~J1655--40, we consider RXTE programs 10255 and 20402, which constitute almost
all of the RXTE exposures of this source.  Spectral parameters for GRO~J1655--40 are
given in Table 1 of Sobczak et al.~(1999a) and the QPO parameters are given in Table~3
of Remillard et al.~(1999a).  The observations of XTE~J1550--564 include exposures
under RXTE programs 30188-06, 30191, 30435, and 40401, with spectral parameters
published in Table~1 of Sobczak et al.~(1999bc) and QPO parameters listed here in
Table~1.  

For four observations listed in Table~1, more than one QPO frequency is given
(observations 153--154 \& 166--167).  In these cases, the QPOs are of approximately
equal amplitude and are blended together.  Consequently, we are unable to identify the
fundamental component, and we have excluded these data from the figures presented
here.  The QPO frequencies of observations 15--75 listed in Table~1 differ slightly
from the corresponding entries (1--60) in Table~1 of Sobczak et al.~(1999b), because
we used the improved fitting technique described above.

\section{Results}

We now investigate the correlations between the frequency/amplitude of the QPOs and
the X-ray spectral parameters of GRO~J1655--40 and XTE~J1550--564. As noted above, our
interest is focused on the strong QPOs in the range 0.08--22~Hz that routinely exhibit
changes in frequency from day to day.  The correlations between QPO frequency and the
spectral parameters are shown in Figures~2 \& 3 and the correlations between QPO
amplitude and the spectral parameters are shown in Figures~4 \& 5.  The open circles
in Figures~2--7 represent those observations in which high-frequency (161--300~Hz)
QPOs are also present.  

For both sources, the variable QPO frequency is correlated with all of the spectral
parameters plotted in Figures 2a--2f \& 3a--3f with a few exceptions (e.g. the
vertical branch with $\nu > 17$~Hz in Fig.~2d).  However, the respective correlations
for the two sources are generally in the opposite sense, although there is a slight
similarity between the QPO frequency vs. $T_{in}$ relation for both sources.  The only
striking exception to the contrary correlations between QPO frequency and the spectral
parameters for the two sources is the relation between QPO frequency and disk flux
shown in Figures~2c \& 2f.  In this case, QPO frequency is correlated with the
temperature and inner disk radius of the disk in such a way that both sources
exhibit a general increase in the QPO frequency as the disk flux increases.  This
relation for XTE~J1550--564 appears roughly linear in the range 1--7~Hz if one
excludes the observations when high-frequency QPOs are present (open circles in
Fig.~2c).  

The QPO amplitude is generally not well correlated with the spectral parameters for
both sources (Fig. 4a--4f \& 5a--5f), and these correlations are more complicated than
the ones considered above.  For XTE~J1550--564, the QPO amplitude is not well
correlated with the temperature or radius of the inner disk (Fig.~4a \& 4b), but the
QPO amplitude does generally decrease as the disk flux increases (Fig.~4c).  Also for
XTE~J1550--564, the QPO amplitude is correlated with the photon index for $\Gamma =
1.5-2.1$, but not well correlated for $\Gamma > 2.1$ (Fig.~5a).  In the case of
GRO~J1655--40, the correlations exhibit two branches, with high-frequency QPOs present
on one branch and absent on the other (Fig. 4d--4f \& Fig.  5d--5f).  For
GRO~J1655-40, note that the branch without high-frequency QPOs consistently has the
steeper slope.  The correlations involving the photon index, power-law flux, and total
flux for XTE~J1550--564 also exhibit two- or possibly three-branches (Fig.~5a--5c). 
However, the branches are less distinct and the high-frequency QPOs are not restricted
to a single branch.  

The correlations between QPO amplitude and QPO frequency are shown in Figure 6a--6b
for both sources.  The QPO amplitude generally decreases as the frequency increases
for both sources.  The QPO amplitude for XTE~J1550--564 is very high and typically
about an order of magnitude greater than the amplitude for GRO~J1655--40.  On the
other hand, the QPO frequency is about twice as high on average for GRO~J1655--40. 
The black hole candidate GRS~1915$+$105 displays QPOs in approximately the same
frequency range and with approximately the same amplitude as XTE~J1550--564.  The
correlations between QPO frequency and amplitude and the spectral parameters for
GRS~1915$+$105 are similar to those of XTE~J1550--564 (Muno, Morgan, \& Remillard
1999).  The larger amplitudes and lower frequencies of the QPOs in XTE~J1550--564 and
GRS~1915$+$105 may indicate that the QPO mechanism in these two sources is operating
in a different regime than for GRO~J1655--40 (due to differences in the mass accretion
rate or central mass) or that a different QPO mechanism is at work.

\section{Discussion}

Both XTE~J1550--564 and GRO~J1655--40 exhibit a similar, strong positive correlation
between the QPO frequency and disk flux in X-rays.  A similar quasi-linear relation
between these two quantities was found for the variable 1--15~Hz QPOs in
GRS~1915$+$105 (Markwardt, Swank, \& Taam 1999) and the 20--30~Hz QPOs in
XTE~J1748--288 (Revnivtsev, Trudolyubov, \& Borozdon 1999).  The correlation between
QPO frequency and disk flux suggests that the QPO frequency is intimately related to
the accretion disk.  Since the contribution of the disk to the total flux in
XTE~J1550--564 and GRO~J1655--40 varies from 5--95\% (Sobczak et al.~1999abc), the
disk flux is not necessarily a good indicator of the total mass accretion rate. 
However, the disk flux alone is a reasonable measure of the mass accretion rate
through the inner disk.  {\it Therefore, we can conclude that the QPO frequency
increases as the mass accretion rate through the inner disk increases.}  

The QPOs are also related to the power-law component: The variable frequency QPOs
appear in XTE~J1550--564 and GRO~J1655--40 only when the power-law contributes more
than 20\% of the 2--20~keV flux (Sobczak et al.~1999abc).  Figure~7 shows the
power-law flux vs. the disk flux.  There is a clear distinction in the observations
with and without QPOs: The observations {\it without} QPOs cluster in a nearly
horizontal line at low power-law flux (Fig.~7b,d), whereas the observations {\it with}
QPOs show significant vertical diplacements toward increased power-law flux
(Fig.~7ac).  The same distinction between observations with and without QPOs is
present in GRS~1915$+$105 (Muno et al.~1999).  {\it We conclude that not only is the
QPO frequency tied to the accretion disk (see above), but in addition the QPO
phenomenon is closely related to the power-law component:  The power-law flux must
reach a threshold of $\sim20$\% of the 2--20~keV flux to trigger the QPO mechanism.  }

The variable frequency QPOs in GRO~J1655--40 and XTE~J1550--564 generally yield rms
amplitudes in the range of 0.3--1.5\% and 1--15\% of the mean count rate,
respectively.  The X-ray luminosity of XTE~J1550--564, which is $\sim 2 \times 10^{38}
$($D$/6kpc)$^2$ erg s$^{-1}$ at typical brightness levels ($\sim$ 2~Crab) when QPOs
are seen, is modulated with a crest-to-trough ratio as high as 1.5 (see Fig.~1b). 
Previously, only the microquasar GRS1915+105 has shown QPOs with such a large
amplitude and a high luminosity (Morgan et al.~1997).  The large modulation of the
X-ray luminosity indicates that these QPOs originate within several gravitational
radii ($r_g = GM/c^2$) of the central object where most of the gravitational energy is
liberated in the accretion flow.  The QPO frequency cannot be the Keplerian frequency
at these radii without some novel mechanism for transporting energy out to large
radii, because a 10~Hz QPO would correspond to a radius of 700~km (or $\sim50~r_g$)
for a $10~M_{\odot}$ black hole.  

As discussed by Markwardt et al. (1999), neither the beat frequency model (Alpar \&
Shaham 1985) nor the sonic point model (Miller, Lamb, \& Psaltis 1998) are applicable
to the $\sim$~10~Hz variable frequency QPOs observed in BHCs.  Thermal-viscous
instabilities (Chen \& Taam 1994; Abramowicz, Chen \& Taam 1995) can produce strong
oscillations; however, their 0.02--0.06~Hz frequencies are well below most of the
0.08--22~Hz QPOs discussed here.  Moreover, this model predicts a negative rather than
a postive correlation between the QPO frequency and mass accretion rate.  One model
that does predict large-amplitude QPOs near 10~Hz, which are positively correlated
with the mass accretion rate, invokes an oscillating shock near the inner disk
(Moltoni, Sponholtz, \& Chakrabarti 1996).  However, it is not clear whether such a
shock is present for accretion flows with high specific angular momentum (Narayan,
Kato, \& Honma 1997).

Another possibility is that the variable frequency QPOs are the result of
radiation-driven oscillations in a quasi-spherical accretion flow or corona (Cui
1999).  Fortner, Lamb, \& Miller (1989) modeled radiation-driven radial oscillations
for neutron stars with luminosities near the Eddington luminosity.  The oscillation
frequency is dominated by the inflow time from the outer part of the radial flow,
which corresponds to $\sim10$~Hz if the radial flow begins a few hundred km from the
central object.  The observed increase in the QPO frequency as the mass accretion rate
through the disk increases could be caused by a shrinking of the radial inflow region.
However, in simulations by Fortner et al.~(1989), the luminosity is modulated by only
$\sim1$\% compared to the almost 15\% modulation observed for XTE~J1550--564.

\section{Conclusions}

For XTE~J1550--564 and GRO~J1655--40, the QPO frequency and amplitude are correlated
at some level with all of the spectral parameters, as illustrated in Figures~2--5. 
However, in the case of the QPO frequency, the correlations are generally opposite for
the two sources.  The only exception is the QPO frequency vs. disk flux plotted in
Figures~2c \& 2f.  {\it Both sources exhibit a general increase in the QPO frequency
as the disk flux increases}.  The correlation between QPO frequency and disk flux
suggests that the QPO frequency is intimately related to the accretion disk.  The same
relation was found for variable 1--15~Hz QPOs in GRS~1915$+$105 (Markwardt et
al.~1999) and 20--30~Hz QPOs in XTE~J1748--288 (Revnivtsev et al.~1999).  Since the
disk flux is a reasonable measure of the mass accretion rate through the disk, we can
conclude that the QPO frequency increases as the mass accretion rate through the disk
increases.  In addition, the QPOs are present only when the power-law contributes more
than 20\% of the 2--20~keV flux, which indicates that both the disk and the power-law
components are linked to the QPO phenomenon.  The QPO amplitude for XTE~J1550--564 is
very high and typically about an order of magnitude greater than the amplitude for
GRO~J1655--40, whereas the QPO frequency is about twice as high for GRO~J1655--40. 
This behavior may indicate that the QPO mechanism is operating in a different regime
for each source, which may relate to why the correlations are generally opposite for
the two sources.  

The correlations between the QPO frequency and amplitude and the X-ray spectral
parameters demonstrate that the QPO phenomenon is linked to the overall emission
properties of the source.  No current model for the origin of these QPOs can
adequately explain more than a few of the many correlations presented here.  We hope
that this study will inspire further significant theoretical work on the subject.  

\acknowledgements This work was supported, in part, by NASA grant NAG5-3680 and the
NASA contract to MIT for instruments of RXTE.  Partial support for J.M. and G.S. was
provided by the Smithsonian Institution Scholarly Studies Program.  W.C. would
like to thank Shuang Nan Zhang and Wan Chen for extensive discussions on spectral
modeling and interpretation of the results.

\newpage
\begin{deluxetable}{cccccc}
\scriptsize
\tablewidth{0pt}
\tablenum{1}
\tablecaption{QPO Parameters for XTE~J1550--564\tablenotemark{\dagger}}
\tablehead{
 \colhead{Obs} & \colhead{Date} & \colhead{MJD\tablenotemark{a}} & 
 \colhead{QPO Freq.\tablenotemark{b}} & \colhead{QPO Amp.\tablenotemark{c}} & 
 \colhead{Q\tablenotemark{d}} \cr
 \colhead{\#} & \colhead{(yymmdd)} & \colhead{ } & 
 \colhead{(Hz)} & \colhead{(\% rms)} & 
\colhead{ }
}
\startdata
1&  980907&   51063.70&    0.084&   13.9$\pm$1.5&   59.7\nl
2&  980908&   51064.01&    0.12&   14.2$\pm$0.6&    6.7\nl
3&  980909&   51065.07&    0.29&   14.3$\pm$0.7&   14.1\nl
4&  980909&   51065.34&    0.39&   15.0$\pm$0.5&    9.4\nl
5&  980910&   51066.07&    0.81&   15.7$\pm$0.4&    8.4\nl
6&  980910&   51066.34&    1.0&   15.6$\pm$0.4&    9.4\nl
7&  980911&   51067.27&    1.6&   16.1$\pm$0.4&    8.3\nl
8&  980912&   51068.35&    2.4&   14.3$\pm$0.2&   11.9\nl
9&  980913&   51069.27&    3.4&   12.9$\pm$0.2&   11.4\nl
10&  980914&   51070.13&    3.2&   13.1$\pm$0.2&   12.8\nl
11&  980914&   51070.27&    3.2&   13.0$\pm$0.2&   12.9\nl
12&  980915&   51071.20&    3.7&   13.0$\pm$0.2&   13.8\nl
13&  980915&   51072.00&    2.6&   11.6$\pm$0.8&   22.9\nl
14&  980916&   51072.34&    4.0&   12.9$\pm$0.2&   15.5\nl
15&  980918&   51074.14&    5.7&   9.4$\pm$0.3&   8.6\nl
16&  980919&   51075.99&   13.1\tablenotemark{e}&   1.05$\pm$0.05&   2.8\nl
17&  980920&   51076.80&    7.2\tablenotemark{e}&   6.6$\pm$0.1&   5.0\nl
18&  980920&   51076.95&    8.5\tablenotemark{e}&   4.8$\pm$0.1&   4.3\nl
19&  980921&   51077.14&    9.8&   3.3$\pm$0.1&   9.2\nl
20&  980921&   51077.21&    7.0&   4.0$\pm$0.1&   5.4\nl
21&  980921&   51077.87&    5.8&   4.8$\pm$0.2&  11.6\nl
22&  980922&   51078.13&    5.4&  10.0$\pm$0.2&  10.7\nl
23&  980923&   51079.79&    4.2&  12.9$\pm$0.4&   7.4\nl
24&  980924&   51080.08&    3.9&  12.2$\pm$0.4&  13.6\nl
25&  980925&   51081.06&    2.9&  13.4$\pm$0.4&   9.8\nl
26&  980926&   51082.00&    2.7&  14.0$\pm$0.4&  10.4\nl
27&  980927&   51083.00&    2.6&  14.4$\pm$0.4&   8.9\nl
28&  980928&   51084.34&    2.7&  14.1$\pm$0.5&  11.0\nl
29&  980929&   51085.27&    4.1&  12.6$\pm$0.3&  11.1\nl
30&  980929&   51085.92&    2.9&  14.3$\pm$0.6&   9.5\nl
31&  980929&   51085.99&    3.0&  14.4$\pm$0.6&   6.5\nl
32&  980930&   51086.89&    3.5&  14.0$\pm$0.2&   8.2\nl
33&  981001&   51087.72&    3.4&  14.2$\pm$0.3&   8.9\nl
34&  981002&   51088.01&    3.2&  14.4$\pm$0.4&   8.6\nl
35&  981003&   51089.01&    3.0&  15.0$\pm$0.5&   8.2\nl
36&  981004&   51090.14&    3.9&  13.3$\pm$0.4&  10.2\nl
37&  981004&   51090.70&    3.7&  13.7$\pm$0.4&   9.5\nl
38&  981005&   51091.74&    5.6&  10.0$\pm$0.3&  11.4\nl
39&  981007&   51093.14&    6.5&   7.5$\pm$0.3&  11.1\nl
40&  981008&   51094.14&    4.3&  12.4$\pm$0.4&  11.1\nl
41&  981008&   51094.57&    5.1&  11.3$\pm$0.3&  11.2\nl
42&  981009&   51095.61&    4.5&  12.1$\pm$0.6&  12.8\nl
43&  981010&   51096.57&    5.5&  11.7$\pm$0.2&   4.9\nl
44&  981011&   51097.57&    4.7&  12.0$\pm$0.5&  10.3\nl
45&  981011&   51097.81&    4.2&  13.5$\pm$0.6&   9.9\nl
46&  981012&   51098.28&    5.0&  11.6$\pm$0.4&  10.1\nl
47&  981013&   51099.22&    4.8&  11.5$\pm$0.5&  11.1\nl
48&  981013&   51099.61&    5.0&  11.7$\pm$0.3&   9.0\nl
49&  981014&   51100.29&    6.5&   7.6$\pm$0.3&  10.9\nl
50&  981015&   51101.61&    6.8&   7.0$\pm$0.3&   6.7\nl
51&  981015&   51101.94&    6.7&   7.0$\pm$0.3&   9.1\nl
52&  981020&   51106.95&    5.5\tablenotemark{e}&   3.8$\pm$0.1&   8.9\nl
53&  981022&   51108.08&    5.4\tablenotemark{e}&   4.1$\pm$0.0&   8.7\nl
54&  981023&   51109.74&    4.9\tablenotemark{e}&   4.0$\pm$0.1&  11.2\nl
59&  981029&   51115.28&    6.8\tablenotemark{e}&   2.8$\pm$0.2&   1.8\nl
64&  981109&   51126.59&    4.9&   2.7$\pm$0.2&   1.9\nl
151&  990302&   51239.08&   18.1&   0.52$\pm$0.06&  17.6\nl
153&  990304&   51241.83&    5.9/10.3$\pm$0.2\tablenotemark{e}&   2.9$\pm$0.1/2.8$\pm$0.1&   3.0/1.6\nl
154&  990305&   51242.51&    5.7/10.0$\pm$0.3\tablenotemark{e}&   2.4$\pm$0.1/2.8$\pm$0.2&   2.5/1.4\nl
155&  990307&   51244.50&   8.4$\pm$0.3\tablenotemark{e}&   1.5$\pm$0.2&   1.8\nl
156&  990308&   51245.35&    6.4\tablenotemark{e}&   3.6$\pm$0.1&  11.7\nl
157&  990309&   51246.41&   7.9$\pm$0.2\tablenotemark{e}&   1.8$\pm$0.1&   1.3\nl
158&  990310&   51247.98&    6.1\tablenotemark{e}&   3.7$\pm$0.1&   9.9\nl
159&  990311&   51248.09&    5.9\tablenotemark{e}&   3.7$\pm$0.1&  11.0\nl
160&  990312&   51249.40&    6.2\tablenotemark{e}&   3.8$\pm$0.1&   9.4\nl
161&  990313&   51250.69&    6.7&   6.5$\pm$0.3&   8.8\nl
162&  990316&   51253.23&    5.5\tablenotemark{e}&   4.3$\pm$0.1&   9.6\nl
163&  990317&   51254.09&    6.1\tablenotemark{e}&   3.8$\pm$0.1&   3.4\nl
164&  990318&   51255.09&    6.2\tablenotemark{e}&   2.4$\pm$0.2&   1.9\nl
166&  990321&   51258.09&    5.2/9.1$\pm$0.9\tablenotemark{e}&   2.6$\pm$0.4/2.8$\pm$0.6&   3.0/1.5\nl
167&  990321&   51258.50&    5.5/10.1$\pm$0.3\tablenotemark{e}&   2.5$\pm$0.2/2.6$\pm$0.3&   2.4/2.2\nl
177\tablenotemark{f}&  990401&   51269.68&    4.5$\pm$0.2&   1.5$\pm$0.4&   3.7\nl
178\tablenotemark{f}&  990402&   51270.74&    8.9\tablenotemark{e}&   2.9$\pm$0.2&   3.5\nl
179\tablenotemark{f}&  990403&   51271.41&    9.1\tablenotemark{e}&   2.8$\pm$0.2&   4.8\nl

\enddata

\tablenotetext{\dagger}{209 PCA observations are given in Sobczak et
al.~(1999c), but only those observtions that exhibit QPOs are listed in this table.}
\tablenotetext{a}{Start of observation, $MJD=JD-2,400,000.5$.}
\tablenotetext{b}{Errors are given at the 95\% confidence level.  The statistical errors 
in the QPO centroid frequency are $<$~1\%, unless given otherwise.}
\tablenotetext{c}{Errors are given at the 95\% confidence level.}
\tablenotetext{d}{Q = Centroid/FWHM}
\tablenotetext{e}{Indicates observations during which high frequency (161--283~Hz)
QPOs are also observed.}
\tablenotetext{f}{PCA Gain Epoch 4 -- The spectral information for
these observations is not as reliable as that obtained for PCA Gain Epoch~3;
therefore, the Epoch~4 data is not included in the figures presented here.}

\end{deluxetable}

\newpage
\begin{figure}
\figurenum{1}
\plotone{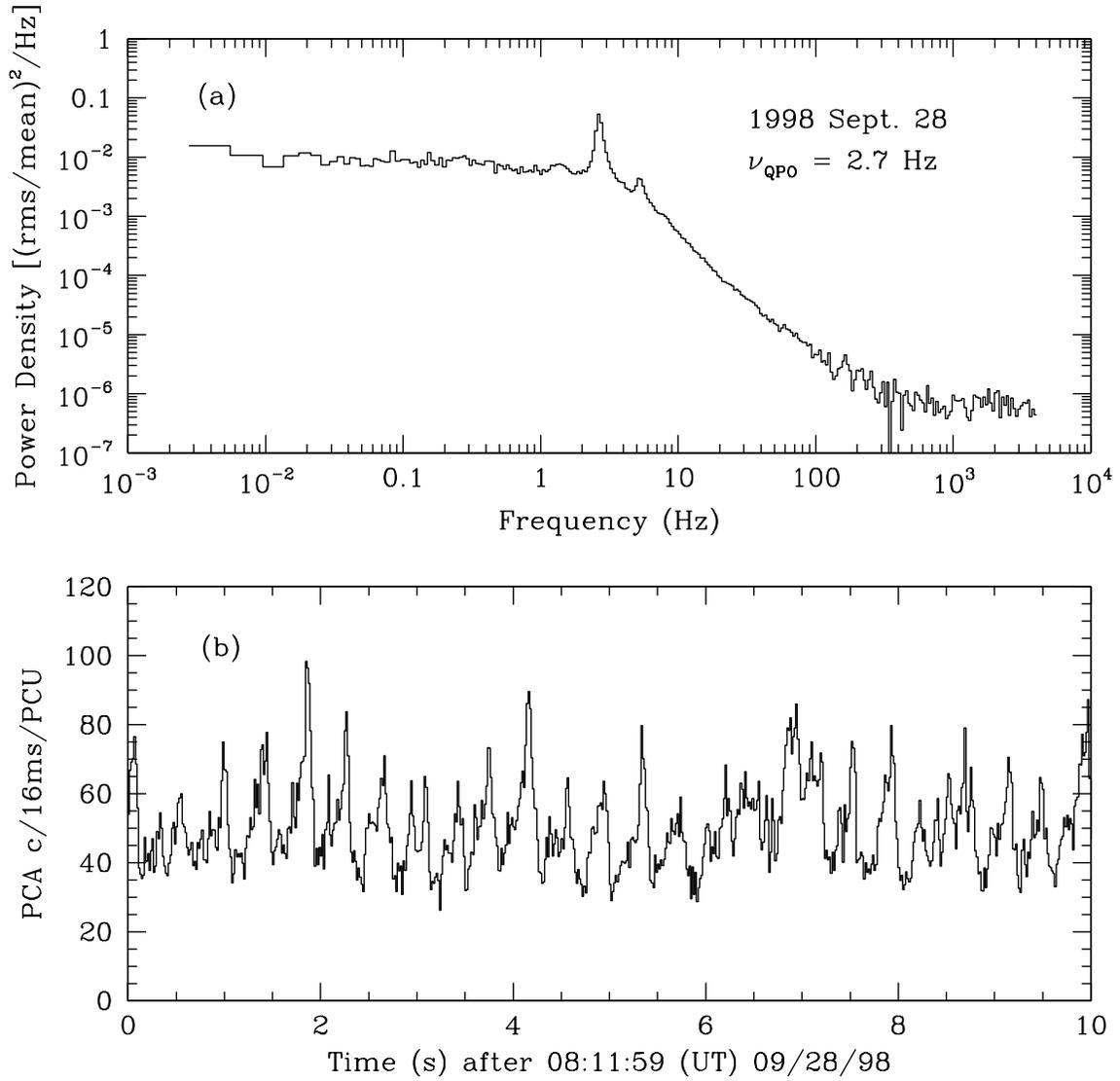}
\caption{Sample power density spectrum and lightcurve of XTE~J1550--564.}
\end{figure}

\newpage
\begin{figure}
\figurenum{2}
\plotone{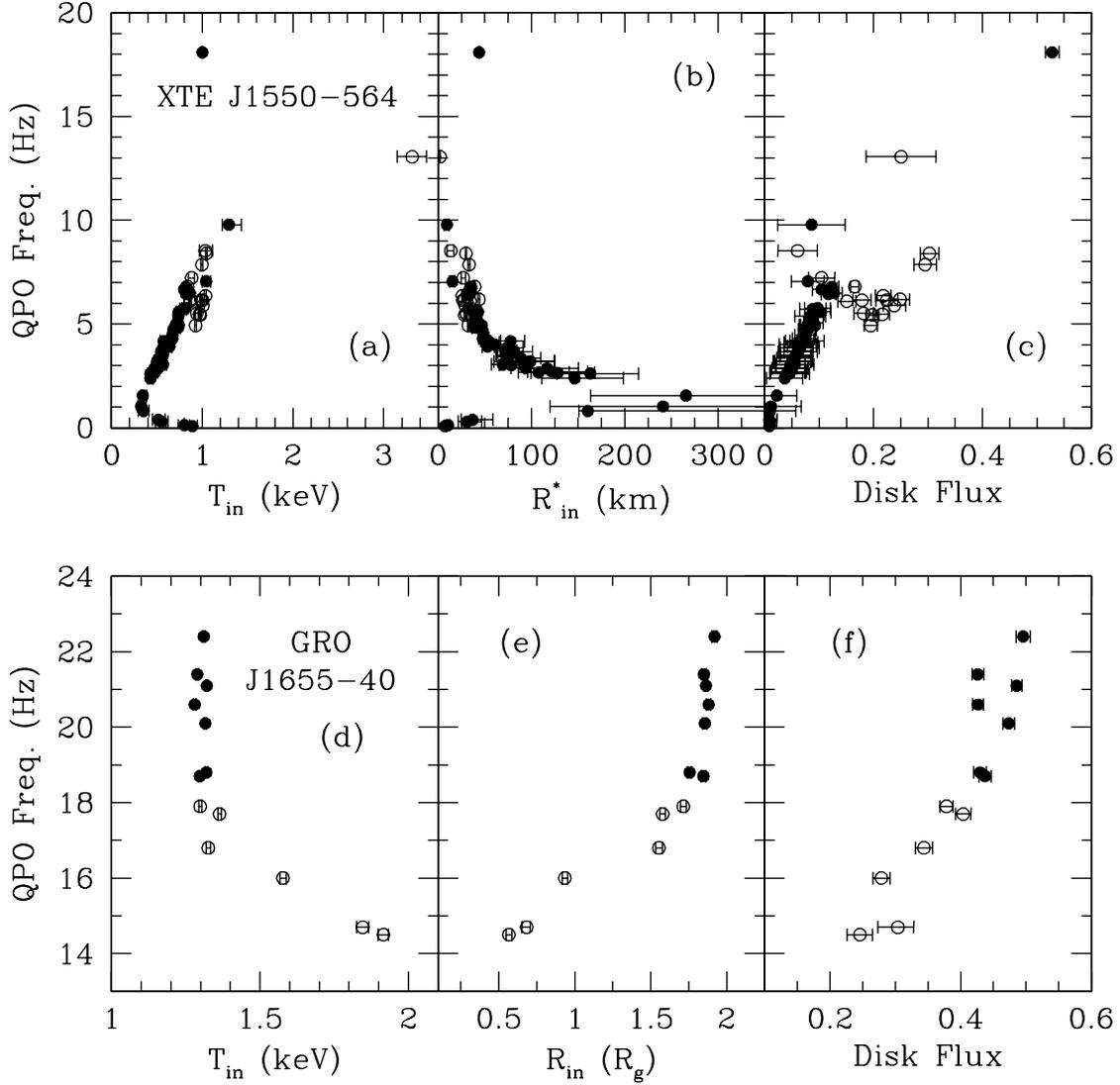}
\caption{{\small QPO frequency vs.~Disk Parameters and Flux for the variable frequency
0.08--22~Hz QPOs in (a-c) XTE~J1550--564 and (d-f) GRO~J1655--40.  The quantities
plotted here are the color temperature of the accretion disk ($T_{in}$) in keV, the
inner disk radius ($R_{in}$), and the unabsorbed 2--20~keV disk flux in units of
$10^{-7}$~erg s$^{-1}$ cm$^{-2}$.  For XTE~J1550--564,
$R^*_{in}=R_{in}(\cos~i)^{1/2}/(D/6~kpc)$ in km, where $i$ is the inclination
angle and $D$ is the distance to the source in kpc.  For GRO~J1655--40, $R_{in}$ is in
units of $GM/c^2=10.4$~km, where $M=7~M_{\odot}$, $D=3.2$~kpc (Hjellming \& Rupen
1995), and $i=69\fdg5$ (Orosz \& Bailyn 1997).  The open symbols represent
observations in which a high-frequency (161--300~Hz) QPO is also observed.  When error
bars are not visible, it is because they are comparable to or smaller than the
plotting symbol.}}
\end{figure}

\newpage
\begin{figure}
\figurenum{3}
\plotone{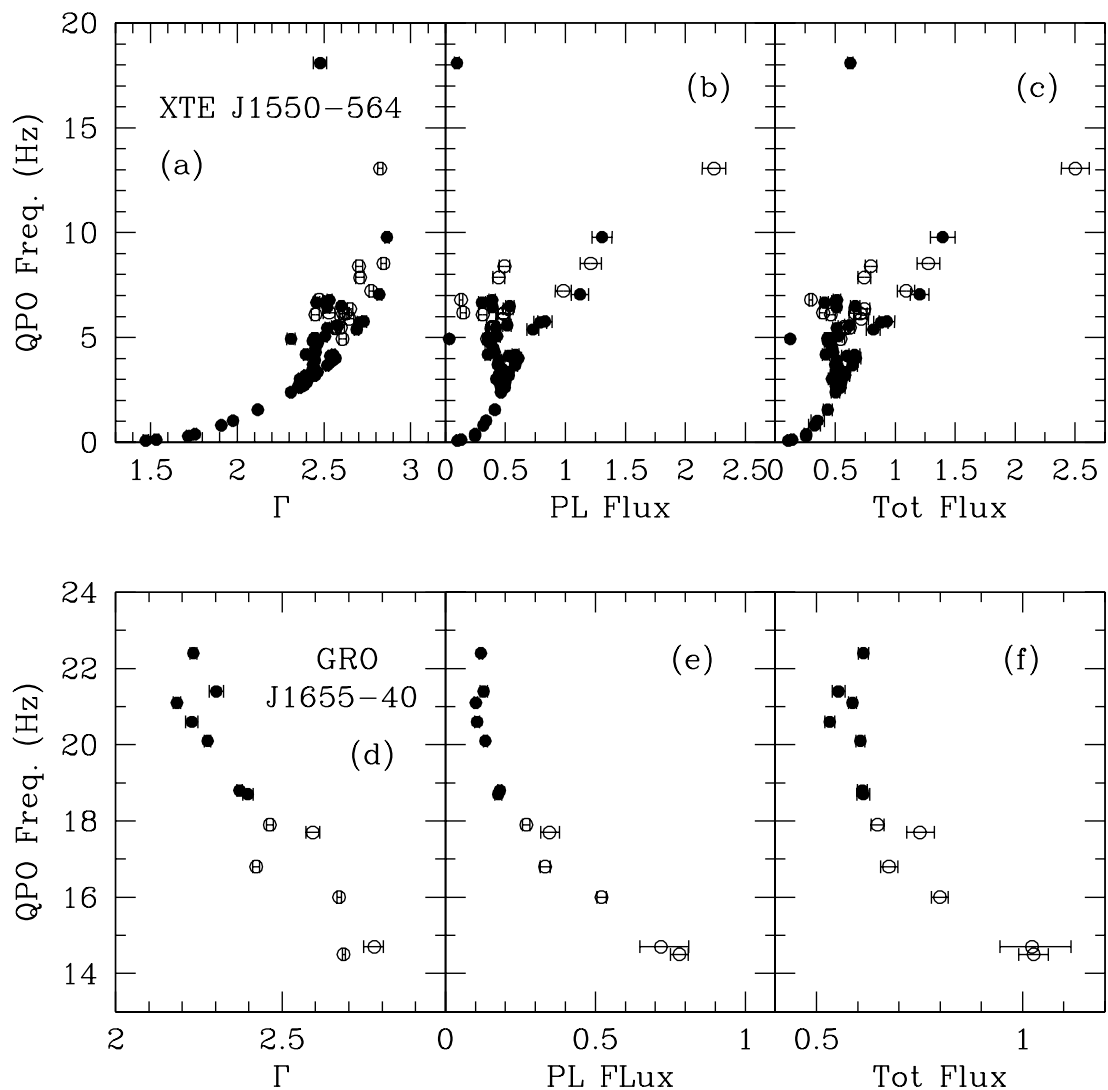}
\caption{QPO frequency vs.~Power-law Parameters and Total Flux for the
variable frequency 0.08--22~Hz QPOs in (a-c) XTE~J1550--564 and (d-f) GRO~J1655--40. 
The quantities plotted here are the power-law photon index $\Gamma$ and the unabsorbed
2--20~keV power-law and total fluxes in units of $10^{-7}$~erg~s$^{-1}$~cm$^{-2}$.}
\end{figure}

\newpage
\begin{figure}
\figurenum{4}
\plotone{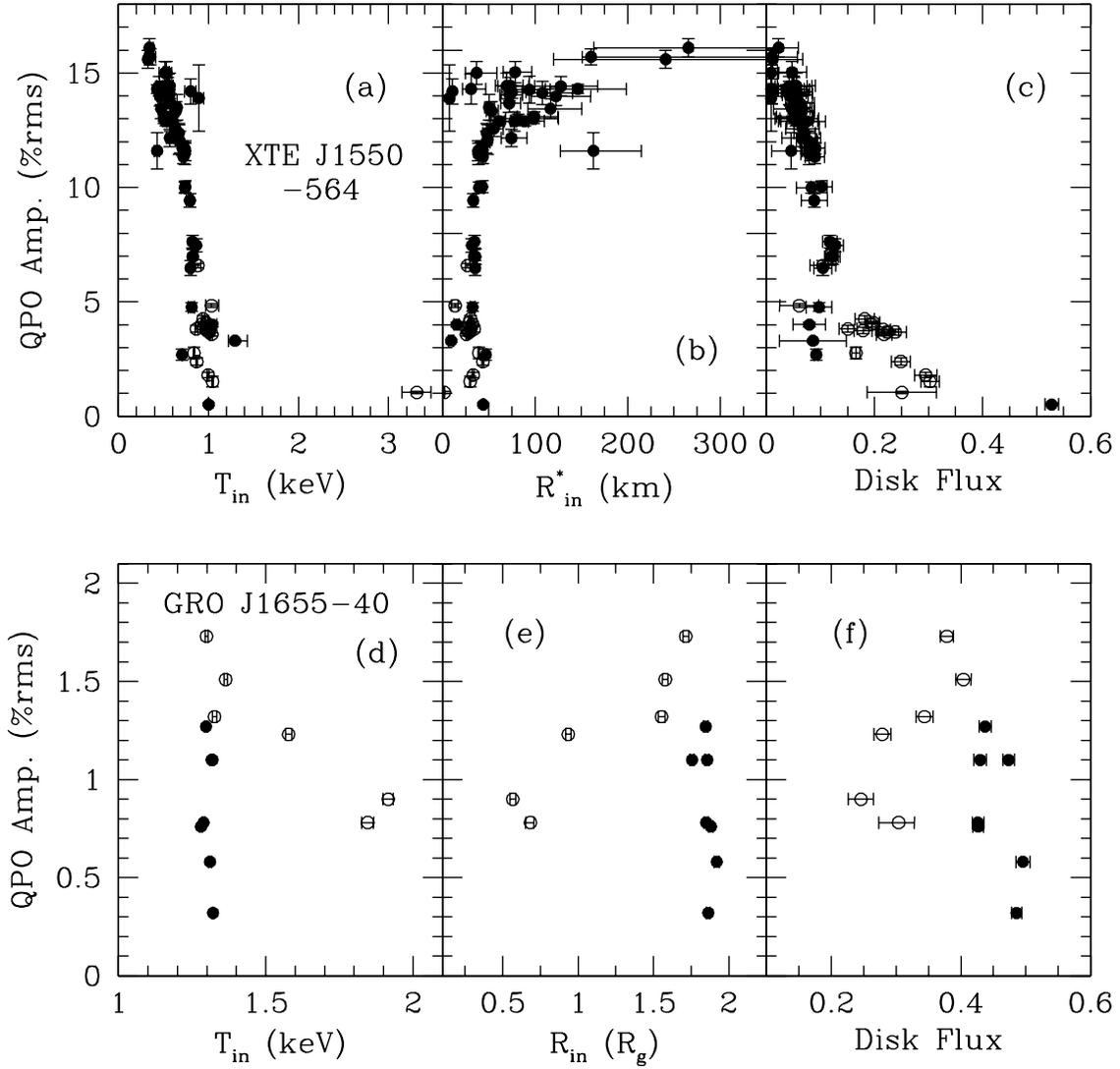}
\caption{QPO Integrated Fractional RMS Amplitude vs.~Disk Parameters and
Flux for the variable frequency 0.08--22~Hz QPOs in (a-c) XTE~J1550--564 and (d-f)
GRO~J1655--40.}
\end{figure}

\newpage
\begin{figure}
\figurenum{5}
\plotone{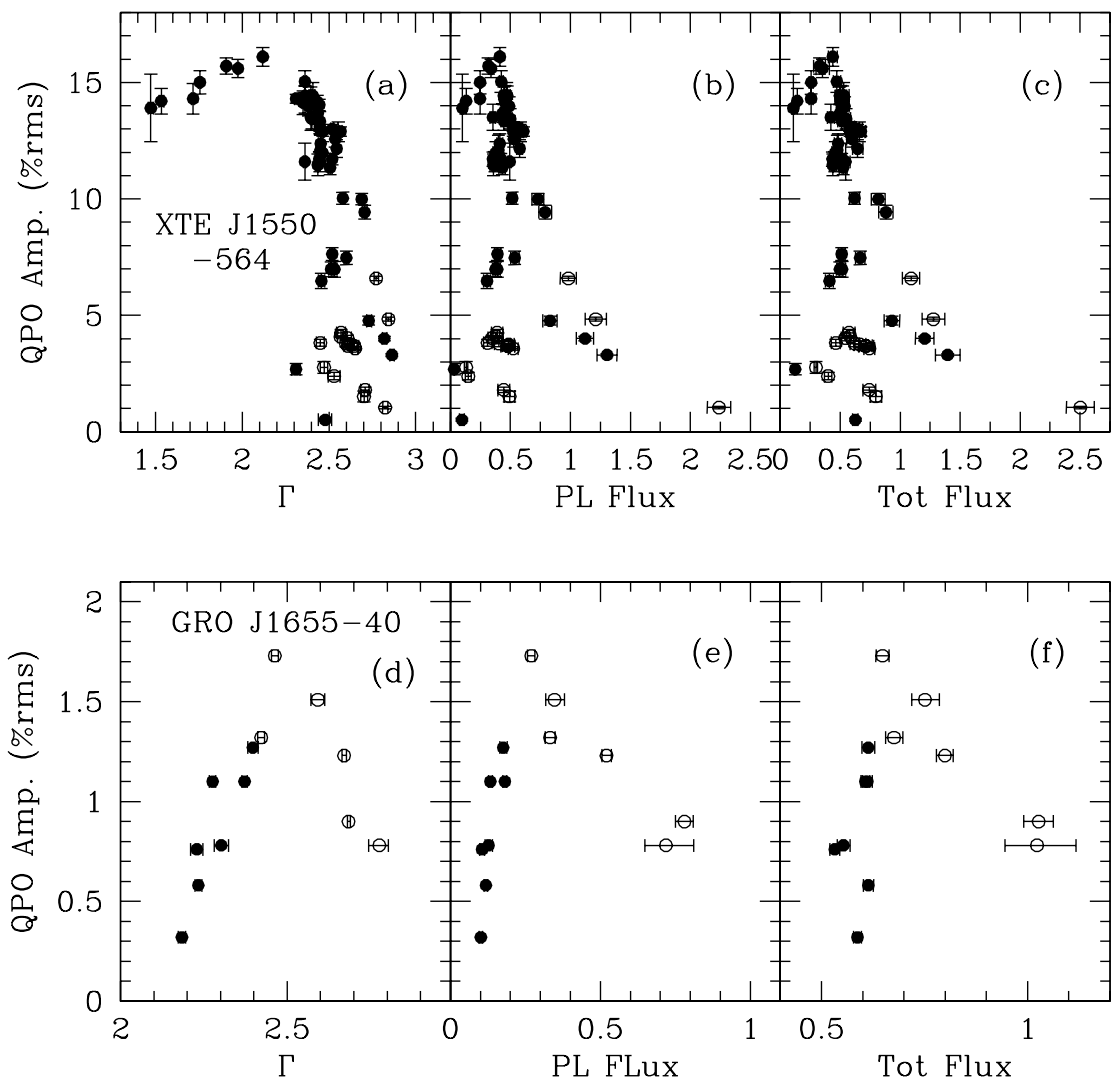}
\caption{QPO Integrated Fractional RMS Amplitude vs.~Power-law Parameters
and Total Flux for the variable frequency 0.08--22~Hz QPOs in (a-c) XTE~J1550--564 and
(d-f) GRO~J1655--40.}
\end{figure}

\newpage
\begin{figure}
\figurenum{6}
\plotone{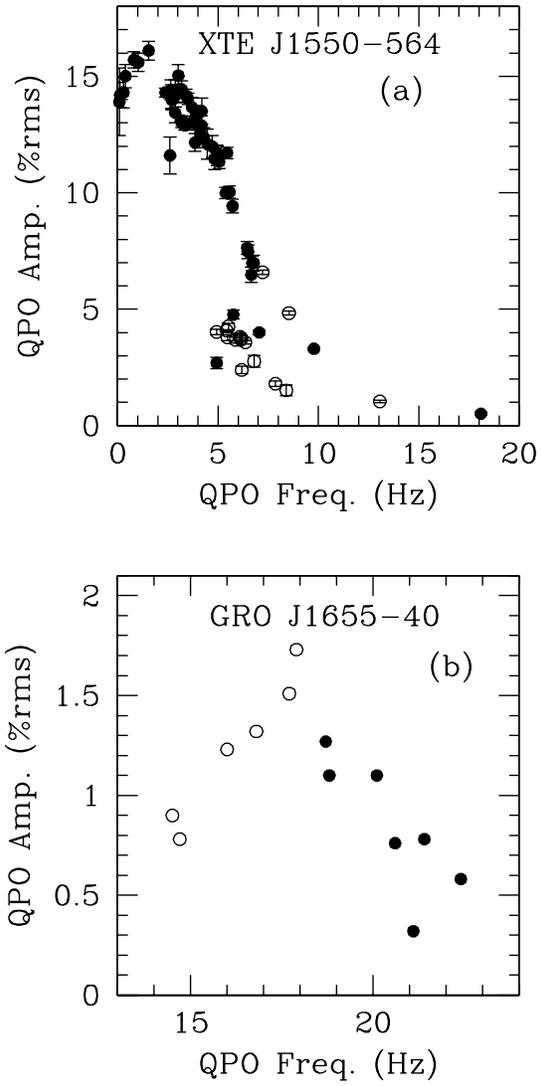}
\caption{QPO Integrated Fractional RMS Amplitude vs.~Frequency for the
variable frequency 0.08--22~Hz QPOs in (a) XTE~J1550--564 and (b) GRO~J1655--40.}
\end{figure}

\newpage
\begin{figure}
\figurenum{7}
\plotone{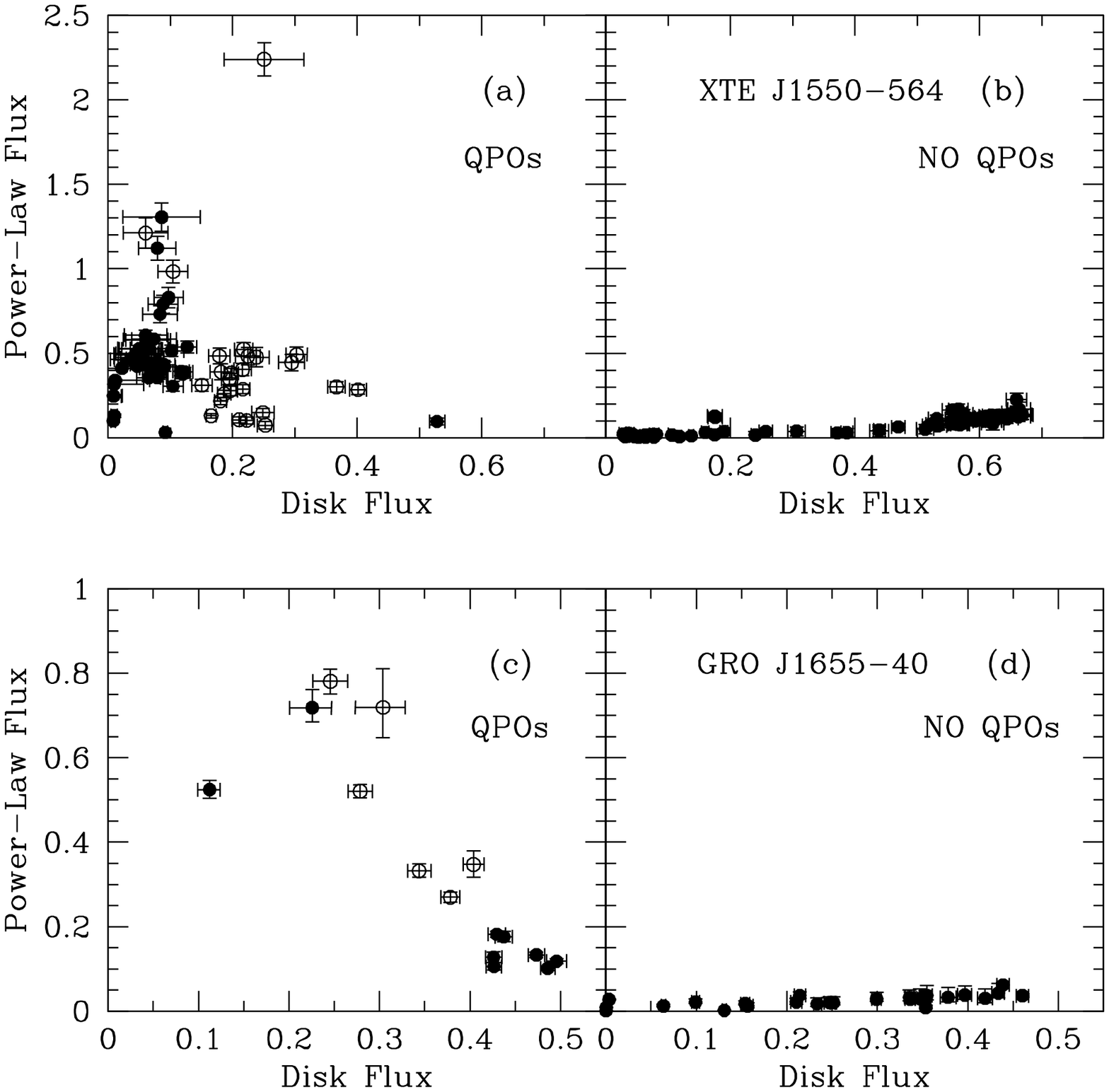}
\caption{Power-Law Flux vs.~Disk Flux for (a,b) XTE~J1550--564 and (c,d)
GRO~J1655--40 when the QPOs are present (a,c) and absent (b,d).  Flux is in units of
$10^{-7}$~erg~s$^{-1}$~cm$^{-2}$.  The open symbols represent observations in which a
high-frequency (161--300~Hz) QPO is also observed.  The two data points in (c) at high
power-law flux for which no high-frequency QPOs are observed are not included in the
previous figures or the discussion of variable frequency QPOs because this QPO near
10~Hz is present in 12 observations with an approximately constant frequency.}
\end{figure}


\begin{references}
\reference{}Abramowicz, M. A., Chen, X., \& Taam, R. E. 1995, \apj, 452, 379
\reference{}Alpar, M. A. \& Shaham, J. 1985, \nat, 316, 239
\reference{}Campbell-Wilson, D., McIntyre, V., Hunstead, R., \& Green, A. 1998,
\iaucirc~7010
\reference{}Chen, X. \& Taam, R. E. 1995, \apj, 441, 354
\reference{}Cui, W. 1999, Proc. ``High Energy Processes in Accreting Black Holes'', 
Eds. J. Poutanen \& R. Svensson (ASP: San Francisco), ASP Conf. Ser. Vol. 161, p. 97 
\reference{}Cui, W., Zhang, S. N., Chen, W., \& Morgan, E. H. 1999, \apj, 512, L43
\reference{}Ebisawa, K., Ogawa, M., Aoki, T., Dotani, T., Takizawa, M., Tanaka, Y.,
Yoshida, K., Miyamoto, S., Iga, S., Hayashida, K., Kitamoto, S., \& Terada, K. 1994, 
\pasj, 46, 375
\reference{}Fortner, B., Lamb, F. K., \& Miller, G. S. 1989, \nat, 342, 775
\reference{}Hjellming, R. M. \& Rupen, M. P. 1995, \nat, 375, 464
\reference{}Jahoda, K., Swank, J. H., Giles, A. B., Stark, M. J., Strohmayer, T.,
Zhang, W., \& Morgan, E. H. 1996, Proc. SPIE 2808, ``EUV and Gamma Ray Instumentation
for Astronomy'' VII, 59
\reference{}Jain, R., Bailyn, C. D., Orosz, J. A., Remillard R. A., \& 
McClintock, J. E. 1999, \apj, 517, L131
\reference{}Kalogera, V. \& Baym, G. 1996, \apj, 470, L61
\reference{}Makishima, K., Maejima, Y., Mitsuda, K., Bradt, H. V., Remillard, R.
A., Tuohy, I. R., Hoshi, R., \& Nakagawa, M. 1986, \apj, 308, 635
\reference{}Markwardt, C. B., Swank, J. H., \& Taam, R. E. 1999, \apj, 513, L37
\reference{}McClintock, J. E. 1998, in Accretion Processes in Astrophysical Systems,
ed. S. S. Holt \& T. Kallman (Woodbury, NY:  AIP), 290
\reference{}Mendez, M., Belloni, T., \& van der Klis, M. 1998, \apj, 499, L187
\reference{}Miller, M. C., Lamb, F. K., \& Psaltis, D. 1998, \apj, 508, 791
\reference{}Mitsuda, K., et al.~1984, PASJ, 36, 741
\reference{}Moltoni, D., Sponholz, H., \& Chakrabarti, S. K. 1996, \apj, 457, 805
\reference{}Morgan, E. H., Remillard, R. A. \& Greiner, J. 1997, \apj, 482, 993
\reference{}Muno, M. P., Morgan, E. H., \& Remillard, R. A. 1999, \apj, submitted,
astro-ph/9904087
\reference{}Narayan, R., Kato, S., \& Honma, F. 1997, \apj, 476, 49
\reference{}Orosz, J. A. \& Bailyn, C. D. 1997, \apj, 477, 876
\reference{}Remillard, R. A., Morgan, E. H., McClintock, J. E., Bailyn, C. D., \& 
Orosz, J. A. 1999a, \apj, 522, 397
\reference{}Remillard, R. A., McClintock, J. E., Sobczak, G. J., Bailyn, C. D., Orosz,
J. A., Morgan, E. H., \& Levine, A. M. 1999b, \apj, 517, L127
\reference{}Revnivtsev, M. G., Trudolyubov, \& Borozdin, K. N. 1999, astro-ph/9903306
\reference{}Rhoades, C. E. \& Ruffini, R. 1974, \prl, 32, 324
\reference{}Sanchez-Fernandez, C. et al.~1999, \aap, 348, L9
\reference{}Shahbaz, T., van der Hooft, F., Casares, J., Charles, P. A., \& van
Paradijs, J. 1999, \mnras, 306, 89
\reference{}Shakura, N. I. \& Sunyaev, R. A. 1973, \aap, 24, 337
\reference{}Shimura, T. \& Takahara, F. 1995, \apj, 445, 780
\reference{}Smith, D. A. \& RXTE/ASM teams 1998, \iaucirc~7008
\reference{}Sobczak, G. J., McClintock, J. E., Remillard, R. A., Bailyn, C. D., \&
Orosz, J. A. 1999a, \apj, 520, 776
\reference{}Sobczak, G. J., McClintock, J. E., Remillard, R. A., Levine, A. M., 
Morgan, E. H., Bailyn, C. D., \& Orosz, J. A. 1999b, \apj, 517, L121
\reference{}Sobczak, G. J., McClintock, J. E., Remillard, R. A., Cui, W., 
Levine, A. M., Morgan, E. H., Orosz, J. A. \& Bailyn, C. D. 1999c, in preparation
\reference{}Tanaka, Y. \& Lewin, W. H. G. 1995, in X-ray Binaries, ed. W. H. G. Lewin,
J. van Paradijs, \& E. P. J. van den Heuvel (Cambridge:  Cambridge Univ. Press), p. 126
\reference{}van der Klis, M. 1995, in X-ray Binaries, ed. W. H. G. Lewin,
J. van Paradijs, \& E. P. J. van den Heuvel (Cambridge:  Cambridge Univ. Press), p. 252
\reference{}Zhang, W., Jahoda, K., Swank, J. H., Morgan, E. H., \& Giles, A. B., 1995,
449, 930
\end{references}
\end{document}